\documentclass[prd,aps,10pt,twocolumn,nofootinbibs,showpacs,reprint]{revtex4}
 
\usepackage[dvips]{graphicx}
\usepackage{amssymb}
\usepackage{amsmath}
\begin{document}
\title{Energy-Momentum Squared Gravity}
\author{Mahmood Roshan$^1$\footnote{mroshan@um.ac.ir}}
\author{Fatimah Shojai$^{2,3}$\footnote{fshojai@ut.ac.ir}}

\affiliation{$^1$Department of Physics, Ferdowsi University of Mashhad, P.O. Box 1436, Mashhad, Iran \\$^2$Department of Physics, University of Tehran, Tehran, Iran\\$^3$Foundations of Physics Group, School of Physics, Institute for Research in Fundamental Sciences (IPM), Tehran, Iran.}

\begin{abstract}
A new covariant generalization of Einstein's general relativity is developed which allows the existence of a term proportional to $T_{\alpha\beta}T^{\alpha\beta}$ in the action functional of the theory ($T_{\alpha\beta}$ is the energy-momentum tensor). Consequently the relevant field equations are different from general relativity only in the presence of matter sources. In the case of a charged black hole, we find exact solutions for the field equations. Applying this theory to a homogeneous and isotropic space-time, we find that there is a maximum energy density $\rho_{\text{max}}$, and correspondingly a minimum length $a_{\text{min}}$, at early universe.  This means that there is a bounce at early times and this theory avoids the existence of an early time singularity. Moreover we show that this theory possesses a true sequence of cosmological eras. Also, we argue that although in the context of the standard cosmological model the cosmological constant $\Lambda$ does not play any important role in the early times and becomes important only after the matter dominated era, in this theory the "repulsive" nature of the cosmological constant plays a crucial role at early times for resolving the singularity.
\end{abstract}
\maketitle
\section{introduction}
 Modifying a gravitational theory dates back to late 1800s, there were some attempts modeled on Maxwell's electrodynamics to modify Newtonian gravity. Since Einstein developed his general relativity (GR) in 1915, various attempts with different motivations have been carried out to generalize it \cite{will}. Some motivations have theoretical character and some observational. Einstein himself modified the original field equations by adding a term including the cosmological constant. Also he proposed the Palatini formulation of GR \cite{ae}. Eddington proposed an interesting alternative to GR in 1924 \cite{edd}. Brans-Dicke theory \cite{bd} and the Einstein-Cartan theory \cite{cartan} are two other examples of a very broad variety of alternatives. Currently, observations of the dark matter and the dark energy provide one of the main motivations for extending GR (for a review on modified gravity theories see e.g. \cite{capo}). 

One of the main intriguing enigmas in GR is that it predicts the existence of space-time singularity at some finite time in the past. However it turns out that GR itself is no longer valid at the singularity because of the expected quantum effects. On the other hand a precise formulation for quantum gravity is still lacking. Nevertheless, there are some classical models in which this kind of singularity can be resolved. For example Eddington-inspired Born-Infeld (EiBI) theory, is a modified theory of gravity which is equivalent to GR only in vacuum and can resolve the singularity \cite{EiBI}. For other examples and also for other motivations behind this kind of modifications, we refer the reader to the review article \cite{bounce}. Here we propose a new model which , despite of its simple appearance, possesses interesting features. Let us start with the following action
\begin{equation}
S=\frac{1}{2\kappa}\int \sqrt{-g}\left(R-2\Lambda-\eta \mathbf{T}^2\right)d^4x+S_M
\label{action}
\end{equation}
where $\mathbf{T}^2=T_{\alpha\beta}T^{\alpha\beta}$, $T_{\alpha\beta}$ is the energy-momentum tensor, $R$ is the Ricci scalar, $\kappa=8\pi G$, $\Lambda$ is the cosmological constant and $S_M$ is the matter action. Also $\eta$ is a coupling constant which its value can be constrained by observations. For a somehow similar approach we refer the reader to \cite{mahmet}. In general $\eta$ can be negative or positive. However as we will show in this paper, a positive $\eta$ leads to a bounce at early universe and to a satisfactory cosmological behavior after the bounce. This bounce avoids the early time singularity. On the other hand, as we will see in section \ref{cosmology}, a negative $\eta$ leads to unsatisfactory cosmological behavior. More specifically there is no stable late time accelerated phase in the case of $\eta<0$. Therefore our main purpose in this paper is to study the $\eta>0$ case.

The situation here is somehow reminiscent of the appearance of the cosmological constant in the standard cosmological model, where $\Lambda$ is postulated to be positive. A negative cosmological constant leads to completely different consequences which are inconsistent with the cosmological observations. More specifically positive $\Lambda$ accelerates the universe while a negative $\Lambda$ decelerates it.

The standard Einstein-Hilbert action can be recovered by setting $\eta=0$. Because of the correction term $\mathbf{T}^2$, we refer to this theory as Energy-Momentum Squared Gravity (EMSG). Throughout the paper, we use units with $c=1$ and assume the metric signature $(-,+,+,+)$ for the metric. It is natural to expect that this correction term be important only in the high energy regimes such as the early universe or within the black holes. Therefore there are no departures from GR in the low curvature regime. 

The outline of the paper is as follows. In section \ref{fe} we derive the field equations of EMSG by varying the action \eqref{action} with respect to the metric. In section \ref{cosmology} we derive the modified Freidmann equations and show that there is a maximum energy density and a minimum length at early universe (when $\eta>0$). Also using the dynamical system method we study the cosmological consequences of EMSG. More specifically we show that this theory possesses a true sequence of cosmological epochs. In section \ref{cbh}, we find an exact charged black hole solution in EMSG. Finally, conclusions are drawn in section \ref{conc}. 

\section{field equations}
\label{fe}
Before moving on to derive the field equations it is important to mention that the correction term $\eta \mathbf{T}^2$ can be defined only when the Lagrangian density for the matter content is specified. Therefore one may not
immediately obtain the field equations from a first-order variation of
the action. In other words, one must first vary the matter action with respect to the gravitational degrees of
freedom. Although this feature is not the case in GR, it appears in theories which introduce correction terms including the energy-momentum tensor in the generic action, for example see \citep{Harko}.

 Comparing the action \eqref{action} with the Einstein-Hilbert action, it is obvious that we need only vary the $\mathbf{T}^2$ term with respect to metric. In other words, the other terms lead to the standard terms in the Einstein field equations. Thus we have
\begin{equation}
\delta(\sqrt{-g}T_{\alpha\beta}T^{\alpha\beta})=\sqrt{-g}\delta(T_{\alpha\beta}T^{\alpha\beta})+T_{\alpha\beta}T^{\alpha\beta}\delta\sqrt{-g}
\label{n1}
\end{equation}
we know that $\delta\sqrt{-g}=-\frac{1}{2}g_{\mu\nu}\sqrt{-g}\delta g^{\mu\nu}$. Therefore the last term in the right hand side of \eqref{n1} can be simply written as
\begin{equation}
-\frac{1}{2}g_{\mu\nu}\mathbf{T}^2\sqrt{-g}\delta g^{\mu\nu}
\end{equation}
Now let us consider the first term in the right hand side of \eqref{n1}. We can write
\begin{equation}
\begin{split}
\delta \mathbf{T}^2&=\delta(g^{\alpha\rho}g^{\beta\sigma}T_{\alpha\beta}T_{\rho\sigma})\\&=2\delta g^{\alpha\rho}T_{\alpha}^{\sigma}T_{\rho\sigma}+2T^{\alpha\beta}\delta T_{\alpha\beta}\\&=2\Big(T_{\mu}^{\sigma}T_{\nu\sigma}+T^{\alpha\beta}\frac{\delta T_{\alpha\beta}}{\delta g^{\mu\nu}}\Big)\delta g^{\mu\nu}
\end{split}
\end{equation}
consequently we obtain
\begin{equation}
\begin{split}
\delta(\sqrt{-g}T_{\alpha\beta}T^{\alpha\beta})=2\Big(&T_{\mu}^{\sigma}T_{\nu\sigma}-\frac{1}{4}g_{\mu\nu}\mathbf{T}^2+\mathbf{\Psi_{\mu\nu}}\Big)\sqrt{-g}\delta g^{\mu\nu}
\end{split}
\end{equation}
where the new tensor $\mathbf{\Psi_{\mu\nu}}$ is defined as
\begin{equation}
\mathbf{\Psi_{\mu\nu}}=T^{\alpha\beta}\frac{\delta T_{\alpha\beta}}{\delta g^{\mu\nu}}
\end{equation}
Finally, bearing in mind that variation of the other terms lead to the convenient terms in the Einstein field equations, the field equations of EMSG can be written as 
\begin{equation}
G_{\mu\nu}+\Lambda g_{\mu\nu}=\kappa T_{\mu\nu}^{\text{eff}}
\label{fe1}
\end{equation}
where $G_{\mu\nu}$ is the Einstein tensor and the effective energy-momentum tensor $T_{\mu\nu}^{\text{eff}}$ is given by
\begin{equation}
T_{\mu\nu}^{\text{eff}}=T_{\mu\nu}+2\frac{\eta}{\kappa}\left(\mathbf{\Psi_{\mu\nu}}+T_{\mu}^{\sigma}T_{\nu\sigma}-\frac{1}{4}g_{\mu\nu}\mathbf{T}^2\right)
\label{n11}
\end{equation}
From \eqref{fe1}, it is clear that $\nabla^{\mu}T_{\mu\nu}^{\text{eff}}=0$ and consequently $\nabla^{\mu}T_{\mu\nu}\neq 0$. We also recall that the matter action $S_M$ can be written as follows
\begin{equation}
S_{M}=\int L_m \sqrt{-g}d^4x
\end{equation}
where $L_m$ is the matter Lagrangian density. The energy-momentum tensor then is defined as
\begin{equation}
T_{\mu\nu}=-\frac{2}{\sqrt{-g}}\frac{\delta (\sqrt{-g}L_m)}{\delta g^{\mu\nu}}
\end{equation}
assuming that $L_m$ depends only on the metric itself and not on its derivatives, we obtain
\begin{equation}
T_{\mu\nu}=g_{\mu\nu}L_m-2\frac{\delta L_m}{\delta g^{\mu\nu}}
\label{n10}
\end{equation}
see \cite{Harko} for more details. As it is clear from field equations \eqref{fe1}, we need the variation of the energy-momentum tensor with respect to the metric. Therefore using equation \eqref{n10} we can write
\begin{equation}
\begin{split}
\frac{\delta T_{\alpha\beta}}{\delta g^{\mu\nu}}&=\frac{\delta g_{\alpha\beta}}{\delta g^{\mu\nu}}L_m+g_{\alpha\beta}\frac{\delta L_m}{\delta g^{\mu\nu}}-2\frac{\partial^2 L_m}{\partial g^{\mu\nu}g^{\alpha\beta}}\\&=-g_{\alpha\mu}g_{\beta\nu}L_m+\frac{1}{2}g_{\alpha\beta}g_{\mu\nu}L_m-\frac{1}{2}g_{\alpha\beta}T_{\mu\nu}\\&~~~~~~-2\frac{\partial^2 L_m}{\partial g^{\mu\nu}g^{\alpha\beta}}
\end{split}
\label{n2}
\end{equation}
where we have also used the following expression
\begin{equation}
\frac{\delta g_{\alpha\beta}}{\delta g^{\mu\nu}}=-g_{\alpha\theta}g_{\beta\rho}\delta^{\theta\rho}_{\mu\nu}
\label{n3}
\end{equation}
where $\delta^{\theta\rho}_{\mu\nu}$ is the generalized Kronecker delta symbol. This relation can be simply obtained using the condition $g_{\alpha\theta}g^{\theta\beta}=\delta^{\beta}_{\alpha}$. Finally multiplying equation \eqref{n2} through by $T^{\alpha\beta}$, we find $\mathbf{\Psi_{\mu\nu}}$ with respect to the matter Lagrangian density
\begin{equation}
\mathbf{\Psi_{\mu\nu}}=-L_m S_{\mu\nu}-\frac{1}{2}T T_{\mu\nu}-2T^{\alpha\beta}\frac{\partial^2 L_m}{\partial g^{\alpha\beta}\partial g^{\mu\nu}}
\label{n4}
\end{equation}
where $S_{\mu\nu}=T_{\mu\nu}-T g_{\mu\nu}/2$, and $T$ is the trace of the energy-momentum tensor. Therefore for a given matter Lagrangian density the field equations \eqref{fe1} are completely known. In the case of a perfect fluid $L_m$ can be simply defined as $L_m=p$ \cite{Harko,barrow}. It is important to mention that the non-relativistic limit of this theory is the same as the Newtonian limit of GR.  Therefore the Poisson equation does not change. Albeit, $\eta$ needs to be small enough to pass the the classical tests of gravity. However, in the presence of matter sources, the higher post-Newtonian corrections will be different from that of GR. 

Before moving on to discuss the cosmology of EMSG, let us first discuss some points about the appearance of $\Lambda$ in the action of EMSG. As we will see in the next section, when $\eta>0$ the correction terms in EMSG are important only in the sufficiently early times, and do not disturb the late time cosmology. Therefore it is evident that without any contribution from other extra fields, such as scalar fields which can enter the matter Lagrangian, EMSG cannot be considered as a dark energy model. Consequently one has to retain the cosmological constant in the theory in order to explain the accelerated expansion of the universe. Of course, as in GR, one may remove the cosmological constant and add new energy contributions to $T_{\alpha\beta}$ in order to construct a dynamical dark energy model. We mean that one can, for instance, add a quintessence scalar field to the theory instead of keeping the cosmological constant.

However, comparing EMSG with GR a question naturally raises. We know that in GR, $\Lambda$ can be a part of the geometric sector. In this case $\Lambda$ is written in the left hand side of the Einstein equations. In this case, let us call it "bare" cosmological constant as it is called in \cite{carroll}. On the other hand it can be a part of the matter action. In this case one may assign an effective perfect fluid energy-momentum tensor to $\Lambda$ with energy density $\rho_{\Lambda}=\Lambda/\kappa$ and pressure $p_{\Lambda}=-\Lambda/\kappa$, i.e. $T_{\mu\nu}^{\Lambda}=p_{\Lambda} g_{\mu\nu}$. In this case $\Lambda$ can be called as "vacuum energy". Albeit the field equations are the same in both pictures.

Then the questions is: Does EMSG lead to different field equations when we put $\Lambda$ in the geometric part or in the matter action? Here we show that EMSG, unlike GR, discriminates between these approaches. This is also the case in other theories which include scalars constructed from the energy-momentum tensor, such as $R_{\mu\nu}T^{\mu\nu}$ and $g_{\mu\nu}T^{\mu\nu}$, in their generic action, for example see \cite{Harko,shahidi}. 

Now let us put the cosmological constant in the matter action. Therefore the total energy-momentum tensor is given by
\begin{equation}
T_{\mu\nu}^{\text{total}}=T_{\mu\nu}+T_{\mu\nu}^{\Lambda}
\end{equation}
and after some algebra, the field equations can be written as 
\begin{equation}
G_{\mu\nu}=\kappa T_{\mu\nu}^{\text{eff}}-\Lambda g_{\mu\nu}+H_{\mu\nu}
\end{equation}
where $T_{\mu\nu}^{\text{eff}}$ is still given by \eqref{n11} and depends only on $T_{\mu\nu}$. All the contributions from $T_{\mu\nu}^{\Lambda}$ have been collected in the new tensor $H_{\mu\nu}$ as
\begin{widetext}
\begin{equation}
H_{\mu\nu}=2\eta\Big(\mathbf{\Psi}_{\mu\nu}^{\Lambda}+T^{\Lambda\,\alpha\beta}\frac{\delta T_{\alpha\beta}}{\delta g^{\mu\nu}}+T^{\alpha\beta}\frac{\delta T_{\alpha\beta}^{\Lambda}}{\delta g^{\mu\nu}}+T_{\mu\sigma}^{\Lambda}T_{\nu}^{\sigma}+T_{\mu}^{\sigma}T_{\nu\sigma}^{\Lambda}+T_{\mu\sigma}^{\Lambda}T_{\nu\theta}^{\Lambda}g^{\sigma\theta}-\frac{1}{4}g_{\mu\nu}(T^{\Lambda}_{\alpha\beta}T^{\Lambda\,\alpha\beta}+2 T^{\Lambda}_{\alpha\beta}T^{\alpha\beta})\Big)
\end{equation}
\end{widetext}
In following we show that $H_{\mu\nu}\neq 0$. This means that by putting the cosmological constant in the matter sector, we get different field equations than \eqref{fe1}. Thus the cosmological behavior, in principle, would be different. Using $T_{\mu\nu}^{\Lambda}=p_{\Lambda} g_{\mu\nu}$ and equations \eqref{n2}-\eqref{n4}, we find
\begin{equation}
\begin{split}
&\mathbf{\Psi}_{\mu\nu}^{\Lambda}=-p_{\Lambda}^2 g_{\mu\nu}\\&
T^{\Lambda\,\alpha\beta}\frac{\delta T_{\alpha\beta}}{\delta g^{\mu\nu}}=-2 p_{\Lambda}T_{\mu\nu}+p_{\Lambda}L_m g_{\mu\nu}\\&
T^{\alpha\beta}\frac{\delta T_{\alpha\beta}^{\Lambda}}{\delta g^{\mu\nu}}=-p_{\Lambda}T_{\mu\nu}\\&
T_{\mu\sigma}^{\Lambda}T_{\nu}^{\sigma}+T_{\mu}^{\sigma}T_{\nu\sigma}^{\Lambda}=2 p_{\Lambda}T_{\mu\nu}\\&
T_{\mu\sigma}^{\Lambda}T_{\nu\theta}^{\Lambda}g^{\sigma\theta}=p_{\Lambda}^2 g_{\mu\nu}\\&
\frac{1}{4}g_{\mu\nu}(T^{\Lambda}_{\alpha\beta}T^{\Lambda\,\alpha\beta}+2 T^{\Lambda}_{\alpha\beta}T^{\alpha\beta})=p_{\Lambda}^2 g_{\mu\nu}+\frac{1}{2} T p_{\Lambda} g_{\mu\nu}
\end{split}
\end{equation}
Therefore $H_{\mu\nu}$ takes the following form
\begin{equation}
H_{\mu\nu}=2\eta \,p_{\Lambda}\Big(\Big(L_m-p_{\Lambda}-\frac{1}{2}T\Big)g_{\mu\nu}-T_{\mu\nu}\Big)
\end{equation}
It is obvious that in this case field equations are more complicated than the case of a bare cosmological constant. More specifically, as expected, $\Lambda^2$ term appears in the field equations. It is worth mentioning that, in GR, an effective cosmological constant can be defined as the linear sum of a bare cosmological constant and the vacuum energy contribution \cite{carroll}. However, as we showed, it is not that simple in EMSG and the gravitational effects of a bare cosmological constant and the vacuum energy cannot be summed simply. On the other hand, in GR, interpreting $\Lambda$ as the vacuum energy raises the so-called cosmological constant problem. Consequently, for the sake of simplicity, we adopt the geometric description, the bare cosmological constant, in this paper. A more general case including the vacuum energy can be considered as a matter of study for future works. Therefore we work with the field equations \eqref{fe1} in what follows.
\section{cosmology of EMSG}
\label{cosmology}
 Let us start with the consequences of this theory in the early universe where we expect significant deviations from the $\Lambda$CDM model. We assume a flat Friedmann-Robertson-Walker (FRW) geometry
\begin{equation}
ds^2=-dt^2+a(t)^2(dx^2+dy^2+dz^2)
\end{equation}
where $a(t)$ is the cosmic scale factor. Also we assume an ideal energy-momentum tensor $T_{\mu\nu}=(\rho+p)u_{\mu}u_{\nu}+pg_{\mu\nu}$ for the cosmic fluid. Using field equation \eqref{fe1}, we find the modified version of the Friedmann equations
\begin{equation}
H^2=\frac{\kappa}{3}\rho+\frac{\Lambda}{3}-\eta\left(\frac{1}{2}p^2+\frac{4}{3}\rho p+\frac{1}{6}\rho^2\right)
\label{fr1}
\end{equation}
\begin{equation}
\frac{\ddot{a}}{a}=-\frac{\kappa}{6}(\rho+3p)+\frac{\Lambda}{3}+\eta\left(p^2+\frac{2}{3}\rho p+\frac{1}{3}\rho^2\right)
\label{fr2}
\end{equation}
where $H=\dot{a}/a$ and the dot denotes a derivative with respect to time. Equations \eqref{fr1} and \eqref{fr2} together with an equation of state form a complete set to study the dynamics of the cosmic fluid and the scale factor. It is worth mentioning that the correction terms in the equations \eqref{fr1} and \eqref{fr2} are somewhat reminiscent of those from quantum geometry effects in loop quantum gravity, for example see \cite{Ashtekar}, or those from the braneworlds \cite{Shtanov:2002mb}. At small energy densities, we recover the standard Friedmann equations. However at high densities a new effect appears: for $\eta>0$, there is a critical point $H=0$ at
\begin{equation}
\rho_{\text{max}}=\frac{\kappa}{\gamma\eta}\left(1+\sqrt{1+\frac{2\eta\Lambda\gamma}{\kappa^2}}\right)
\end{equation}
Where $\gamma=3w^2+8w+1$ and we have assumed a barotropic equation of state $p=w \rho$ ($w>0$). In the very early universe, $\rho_{\text{max}}$ is an explicit cutoff in the energy density. In fact in this era $w=1/3$ and the maximum energy density is $\rho_{\text{max}}=(1+\sqrt{1+8\Lambda\eta/\kappa^2})\kappa/4\eta$. In other words, this means that the early radiation dominated universe does not start from a singularity. More surprisingly, the universe passes across a regular bounce at this point. One may easily verify that at this point $\ddot{a}=\frac{2\Lambda}{3}a>0$. Fortunately, in the radiation dominated universe the field equations \eqref{fr1} and \eqref{fr2} can be exactly solved. The result is
\begin{equation}
\begin{split}
&a(t)=a_{\text{min}}\sqrt{\cosh\alpha t}\\& \rho_r(t)=\frac{\kappa}{4\eta}\left(1+\sqrt{1+\frac{8\eta\Lambda}{\kappa^2}~\text{sech}^2\alpha t}\right)
\end{split}
\end{equation} 
where $\alpha=\sqrt{\frac{4\Lambda}{3}}$. We recall that the corresponding solution in $\Lambda$CDM model is $a(t)\sim \sqrt{\sinh\alpha t}$. The main difference is that unlike the $\Lambda$CDM case, $\ddot{a}$ is positive in EMSG. This is reminiscent of an inflationary epoch without any scalar field. Note that the effective energy density and pressure remain also finite and all effective energy conditions fail at the bounce. The effective energy density and pressure, $\rho_{\text{eff}}$ and $p_{\text{eff}}$, are defined using the effective energy-momentum tensor $T_{\mu\nu}^{\text{eff}}$. At the bounce we have $\rho_{\text{eff}}=-\Lambda/\kappa<0$ and $p_{\text{eff}}=\rho_{\text{eff}}/3$. It is clear that the null energy condition is violated. It is well known that in bouncing cosmologies one or more energy conditions are violated \cite{bounce}. For generalized energy conditions in modified theories of gravity we refer the reader to \cite{capo2}. 

It is clear that in this theory, existence of a positive cosmological constant is necessary for preventing the singularity. In fact the repulsive nature of the cosmological constant plays an important role here. On the other hand, in the $\Lambda$CDM model, $\Lambda$ does not have a significant effect in the early universe and its role is dominant at sufficiently the late times. 

A maximum in energy density naturally implies a minimum value for the cosmic scale factor. Here we estimate this minimum length $a_{\text{min}}$. As we mentioned before, in EMSG the perfect fluid does not satisfy the standard conservation laws. For a two component cosmic fluid, the conservation equation $\nabla^{\mu}T_{\mu\nu}^{\text{eff}}=0$ includes some interaction terms proportional to $\dot{\rho_m}\rho_r$, $\dot{\rho_r}\rho_m$ and $\rho_m\rho_r$, where $\rho_m$ is the matter energy density. One may equally distribute these interaction terms between two energy components. In this case the conservation equation can be split up into two separate equations
 \begin{equation}
\frac{\bar{\rho}_r'}{\bar{\rho}_{{r}}} ={\frac { \left( 9\bar{\rho}_{m}^{2}+48\bar{\rho}_{{r}}\bar{\rho}_{{m}
}+56\bar{\rho}_{{r}}^{2}-66\bar{\rho}_{{m}}-152\bar{\rho}_{{r}}+96 \right) }{26
\bar{\rho}_{{m}}+62\bar{\rho}_{{r}}-7
\bar{\rho}_{{m}}^{2}-24\bar{\rho}_{{r}}\bar{\rho}_{{m}}-28\bar{\rho}_{{r}}^{2}-24}}
\label{rad}
 \end{equation}
 \begin{equation}
\frac{ \bar{\rho}_m'}{\bar{\rho}_{{m}}}={\frac {\left( 21\bar{\rho}_{{m}}^{2}+72\bar{\rho}_{{r}}\bar{\rho}_{{m
}}+64\bar{\rho}_{{r}}^{2}-78\bar{\rho}_{{m}}-148\bar{\rho}_{{r}}+72 \right) }{26
\bar{\rho}_{{m}}+62\bar{\rho}_{{r}}-7
\bar{\rho}_{{m}}^{2}-24\bar{\rho}_{{r}}\bar{\rho}_{{m}}-28\bar{\rho}_{{r}}^{2}-24}}
\label{7}
 \end{equation}
 where $\bar{\rho}_i=\frac{2\eta\rho_i}{\kappa}$ and prime denotes derivative with respect to $\ln a$. For small $\bar{\rho}_i$ we recover the conventional conservation equations. These conservation equations are complicated and so it is not easy to find the exact value of $a_{\text{min}}$. In order to obtain an estimate for $a_{\text{min}}$, we neglect $\bar{\rho}_m$ in the conservation equation of the radiation \eqref{rad}. This assumption makes sense because $\bar{\rho}_m$ is very smaller than $\bar{\rho}_r$ in the beginning and it decreases even more after the bounce. Therefore, although eventually $\bar{\rho}_r$ will fall bellow $\bar{\rho}_m$, during a long period of time compared to the age of the universe this approximation holds. In this case, equation \eqref{rad} can be expressed as
\begin{equation}
\bar{\rho}_r'+4\bar{\rho}_r\simeq-\sum_{n=2}^{\infty}2^n\bar{\rho}_r^n
\end{equation}
by retaining only the dominant term on the right hand side, we find 
\begin{equation}
\bar{\rho}_r=\frac{a_{\text{min}}^4\rho_{\text{max}}}{(\kappa/2\eta+\rho_{\text{max}})a^4-a_{\text{min}}^4\rho_{\text{max}}}
\end{equation}
finally one may simply show that
\begin{equation}
a_{\text{min}}\simeq \left(\frac{12\eta}{\kappa^2}\right)^{1/4}\left(H_0^2 \Omega_{\text{r0}}\right)^{1/4} a_0
\end{equation} 
where $\Omega_{\text{r}}=\frac{\kappa\rho_{r}}{3H^2}$ is the radiation density parameter and subscript $0$ denotes the present value of the quantities. This is the minimum length in the cosmology of EMSG.  
\begin{figure}[!t]
\hspace{0pt}\rotatebox{0}{\resizebox{0.48\textwidth}{!}{\includegraphics{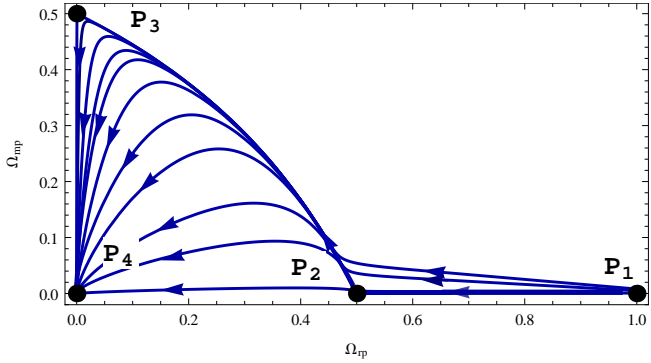}}}
\vspace{-15pt}{\caption{2-dimensional phase space trajectories in the Poincar\'{e} projected phase space. The fixed point $P_1$ is an unstable very early time radiation dominated fixed point appearing in the "infinity" of the system. $P_2$ corresponds to a standard radiation era. The fixed point $P_3$ is a true matter era, and $P_4$ is the final late time de Sitter attractor.  
}\label{1}}
\end{figure}

Although this theory avoids the singularity, it should possess a valid expansion history.  In other words, it must start with a radiation era. Also it has to possess a mater dominated era followed by an accelerated expansion. In principle, the correction terms in the new Friedmann equations can destroy the standard thermal history. In order to check this important requirement and to determine whether EMSG provides a viable alternative to the standard model, we use \textit{dynamical system analysis} \cite{perko}. The advantage of the finite dynamical systems analysis is to provide a fast and numerically reliable integration of the modified Friedmann equations. Using this method, one may link the early and late time evolution of the model by analyzing the fixed points in the compact phase-space of the model, for example see \cite{amendola}. The expansion history can then be easily compared to the standard $\Lambda$CDM model. This method has been applied to several alternative theories and cosmological models, for example see \cite{Xu:2012jf}.

We use cosmic density parameters $\Omega_{\text{m}}=\frac{\kappa\rho_m}{3H^2}$, $\Omega_{\text{r}}$ and $\Omega_{\Lambda}=\frac{\Lambda}{3H^2}$ as phase space variables. Using modified Friedmann equations \eqref{fr1} and \eqref{fr2} and also the corresponding conservation equations for matter and radiation \eqref{rad} and \eqref{7}, one can find three autonomous equations. These equations take the following form
\begin{equation}
\begin{split}
&{\Omega_{\text{m}}'}=f_1(\Omega_\text{m},\Omega_\text{r},\Omega_{\Lambda})\\&
{\Omega_{\text{r}}'}=f_2(\Omega_\text{m},\Omega_\text{r},\Omega_{\Lambda})\\&
{\Omega_{\Lambda}'}=f_3(\Omega_\text{m},\Omega_\text{r},\Omega_{\Lambda})
\end{split}
\label{n12}
\end{equation}
where functions $f_i$ are given by
\begin{widetext}
\begin{equation}
\begin{split}
&f_1(\Omega_\text{m},\Omega_\text{r},\Omega_{\Lambda})={\frac {-21{\Omega_{{\text{m}}}}^{3}+ \left( -72\Omega_{{\text{r}}}+78\Omega_{
{\max}} \right) {\Omega_{{\text{m}}}}^{2}+ \left( -64{\Omega_{{\text{r}}}}^{2}+148
  \Omega_{{\text{r}}}  \Omega_{{\max}}  -72{
\Omega_{{\max}}}^{2} \right)  \Omega_{{\text{m}}}  }{7{\Omega
_{{m}}}^{2}+ \left( 24\Omega_{{\text{r}}}-26\Omega_{{\max}} \right) 
  \Omega_{{\text{m}}}  +28{\Omega_{{\text{r}}}}^{2}-62 \left( 
\Omega_{{\text{r}}} \right)   \Omega_{{\max}}  +24{\Omega_{{
\max}}}^{2}}}\\&~~~~~~~~~~-\frac{1}{2}{\frac { \left( \Omega_{{\text{m}}}+2\Omega_{{\text{r}}}-2
\Omega_{{\max}} \right)  \Omega_{{\text{m}}}  \left( 3\Omega
_{{m}}+4\Omega_{{\text{r}}} \right) }{\Omega_{{\max}}}}\\&
f_2(\Omega_\text{m},\Omega_\text{r},\Omega_{\Lambda})=-{\frac { \Omega_{{\text{r}}}   \left( 56{\Omega_{{\text{r}}}}^{2}+
48  \Omega_{{\text{r}}}    \Omega_{{\text{m}}}  -152
  \Omega_{{\text{r}}}   \Omega_{{\max}} +9{
\Omega_{{\text{m}}}}^{2}-66  \Omega_{{\text{m}}}   \left( \Omega_{{
\max}} \right) +96{\Omega_{{\max}}}^{2} \right) }{7{\Omega_{{\text{m}}}}^
{2}+ \left( 24\Omega_{{\text{r}}}-26\Omega_{{\max}} \right)  \left( 
\Omega_{{\text{m}}} \right) +28{\Omega_{{\text{r}}}}^{2}-62 \left( \Omega_{{\text{r}}}
 \right)  \Omega_{{\max}}  +24{\Omega_{{\max}}}^{2}}}\\&~~~~~~~~~~-
\frac{1}{2}{\frac {  \Omega_{{\text{r}}}   \left( 3\Omega_{{\text{m}}}+4
\Omega_{{\text{r}}} \right)  \left( \Omega_{{\text{m}}}+2\Omega_{{\text{r}}}-2\Omega_{{
\max}} \right) }{\Omega_{{\max}}}}\\&
f_3(\Omega_\text{m},\Omega_\text{r},\Omega_{\Lambda})=-\frac{3}{2}{\frac {\Omega_{{\Lambda}}  \left( \Omega_{{\text{m}}}
+2\Omega_{{\text{r}}}-2\Omega_{{\max}} \right)  \left( \Omega_{{\text{m}}}+4/3
\Omega_{{\text{r}}} \right) }{\Omega_{{\max}}}}
 \end{split}
 \label{at}
\end{equation}
\end{widetext}
where 
\begin{equation}
\Omega_{{\max}}=\frac{\kappa\rho_{\max}}{3H^2}=\frac{1}{12}{\frac {3{\Omega_{{\text{m}}}}^{2}+14 \Omega_{{\text{r}}}  
\Omega_{{\text{m}}}  +12{\Omega_{{\text{r}}}}^{2}}{-1+\Omega_{{
\Lambda}} \left( t \right) +\Omega_{{\text{m}}}+\Omega_{{\text{r}}}}}
\end{equation}
 It is interesting that although the field equations and conservation equations seem complicated, as we will show, the cosmological behavior of the model is close to the $\Lambda$CDM model.

In order to find the critical/fixed points of the dynamical system \eqref{n12}, it is just enough to set to zero functions $f_i$ and find the relevant roots. Then we can determine whether the system approaches one of the
critical points or not by analyzing the stability around the
critical points. Let us consider small perturbations $\delta\Omega_\text{m}$, $\delta\Omega_\text{r}$ and
$\delta\Omega_{\Lambda}$ around the critical point $(\Omega_\text{m}^c,\Omega_\text{r}^c,\Omega_{\Lambda}^c)$, namely
\begin{equation}
\begin{split}
&\Omega_\text{m}=\Omega_\text{m}^c+\delta\Omega_\text{m}\\&
\Omega_\text{r}=\Omega_\text{r}^c+\delta\Omega_\text{r}\\&
\Omega_{\Lambda}=\Omega_{\Lambda}^c+\delta\Omega_{\Lambda}
\end{split}
\end{equation}
Substituting into equations \eqref{at} yields to the linear differential equations
\begin{equation}
\left( \begin{array}{c}
\delta\Omega_\text{m}' \\
\delta\Omega_\text{r}' \\
\delta\Omega_{\Lambda}'\end{array} \right)= \mathfrak{M}\left( \begin{array}{c}
\delta\Omega_\text{m} \\
\delta\Omega_\text{r} \\
\delta\Omega_{\Lambda}\end{array} \right)
\end{equation}
The stability matrix $\mathfrak{M}$ depends upon $\Omega_\text{m}^c$, $\Omega_\text{r}^c$ and $\Omega_{\Lambda}^c$, and is given by
\begin{equation}
\mathfrak{M}={\left( \begin{array}{ccc}
\frac{\partial f_1}{\partial \Omega_{\text{m}}} & \frac{\partial f_1}{\partial \Omega_{\text{r}}} & \frac{\partial f_1}{\partial \Omega_{\Lambda}} \\
\frac{\partial f_2}{\partial \Omega_{\text{m}}} & \frac{\partial f_2}{\partial \Omega_{\text{r}}} & \frac{\partial f_2}{\partial \Omega_{\Lambda}}\\
\frac{\partial f_3}{\partial \Omega_{\text{m}}} & \frac{\partial f_3}{\partial \Omega_{\text{r}}} & \frac{\partial f_3}{\partial \Omega_{\Lambda}} \end{array} \right)_{(\Omega_\text{m}^c,\Omega_\text{r}^c,\Omega_{\Lambda}^c)}}
\end{equation}
for each fixed point, this matrix possesses three eigenvalues. The fixed point is stable/attractor if all the eigenvalues are negative and is unstable if there is at leas one positive eigenvalue.

We have found the critical points and developed the stability matrix. Here are the results: as in the $\Lambda$CDM model there are two repulsive critical points $P_2:(\Omega_{\text{m}}^c,\Omega_{\text{r}}^c,\Omega_{\Lambda}^c)=(0,1,0)$, $P_3=(1,0,0)$. The eigenvalues for theses points are $(1,4,-4)$ and $(-3,3,-1)$ respectively. As it is clear, $P_2$ corresponds to a standard unstable radiation dominated era, and $P_3$ is a true unstable matter dominated point. In fact the effective equation of state parameter $\omega_{\text{eff}}$, see \eqref{n13} for definition, for $P_2$ is equal to $1/3$ (this is also evident from Fig \ref{3}). Therefore the cosmic scale factor grows as $a(t)\propto t^{1/3}$. On the other hand for $P_3$ we have $\omega_{\text{eff}}=0$ and therefore $a(t)\propto t^{2/3}$.

Also there is a late time attractor $P_4=(0,0,1)$. In this case the eigenvalues are $(0,-6/5,-4)$. This point corresponds to a late time de Sitter expansion. Let us recall again that $\Lambda$ has a twofold task in this model, providing a regular bounce and triggering the late time cosmic speed up.
\begin{figure}[!t]
\hspace{0pt}\rotatebox{0}{\resizebox{0.45\textwidth}{!}{\includegraphics{{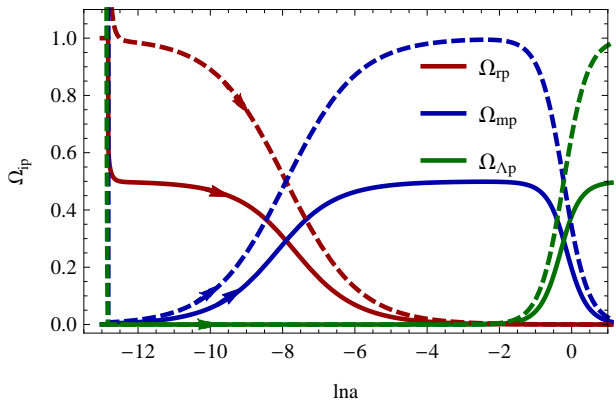}}}}
\vspace{-12pt}{\caption{Behavior of the cosmic density parameters for a specific choice of initial conditions at $\ln a=-12.5$ as $\Omega_{\text{m}}=0.01$, $\Omega_{\text{r}}=1$ and $\Omega_{\Lambda}\simeq 0$. Thick curves show the projected density parameters and the dashed curves correspond to the density parameters themselves.
}\label{2}}
\end{figure}

It is also necessary to check the behavior of this system at "infinity". Note that at the bounce the phase variables become infinite and thus they do not make a compact phase space. In order to demonstrate the main features of the system in a compact region (a sphere with radius $1$) including the infinity, we use the Poincar\'{e} coordinates, obtained by the transformation $\Omega_{\text{ip}}=\Omega_{\text{i}}/(1+\sqrt{\Omega_{\text{m}}^2+\Omega_{\text{r}}^2+\Omega_{\Lambda}^2})$. Using these new variables it becomes clear that there is another unstable radiation dominated fixed point at infinity, i.e. $P_1:(\Omega_{\text{mp}},\Omega_{\text{rp}},\Omega_{\Lambda\text{p}})=(0,1,0)$. In Fig. \ref{1} we show a 2-dimensional phase space plot.  The above mentioned critical points have been shown in this phase plot. Therefore we see that the evolution can be started from $P_1$ and rapidly repels away from it and is followed by a standard radiation dominated epoch $P_2$. Then the models enters the matter era $P_3$ and finally falls into the late time de Sitter attractor $P_4$. In fact after the rapid evolution from $P_1$ to $P_2$, the system follows standard trajectories similar to $\Lambda$CDM model. In Fig. \ref{2} we have plotted the evolution of the cosmic density parameters for a specific choice of initial conditions. As it is clear from this figure, after a rapid decrease in $\Omega_{\text{rp}}$ system evolves as standard cosmological model and the current values of the density parameters can match the observed values. 

In order to emphasize that EMSG with a bare cosmological constant does not disturb the standard cosmic evolution after the bounce, we have plotted the behavior of the deceleration parameter $q$ and the effective equation of state parameter $\omega_{\text{eff}}$ in Fig. \ref{3}. These parameters belong to the numerical solution presented in Fig \ref{2}. We recall that
\begin{equation}
q=-\frac{\ddot{a}a}{\dot{a}^2},~~~~~~~\omega_{\text{eff}}=-1-\frac{2\dot{H}}{3 H^2}
\label{n13}
\end{equation} 
These parameters can be written with respect to the phase space variables. Since the relevant expressions are too long, we have not written them. In Fig. \ref{3}, the solid lines correspond to 
EMSG and the dashed curves correspond to $\Lambda$CDM model. The purple regions demonstrate the approximate time intervals for which the system is close to its critical points. It is worth mentioning that the autonomous equations in $\Lambda$CDM model can be obtained from \eqref{at} by considering the limit $\Omega_{\text{max}}\rightarrow \infty$. In Fig \ref{3}, the initial conditions in $\Lambda$CDM model are chosen in a way to lead the same values as in EMSG for the cosmic density parameters, i.e. $\Omega_i$, at present epoch $\ln a=0$. As it is obvious, $q$ and $\omega_{\text{eff}}$ are exactly the same in both theories. This means that the growth rate of the cosmic scale factor in EMSG coincides with the standard case. Consequently after the bounce the thermal history of the universe is the same. It should be noted that this behavior does not belong only to the above mentioned initial condition, and we have checked it for several initial conditions. 

\begin{figure}[!t]
\hspace{0pt}\rotatebox{0}{\resizebox{0.45\textwidth}{!}{\includegraphics{{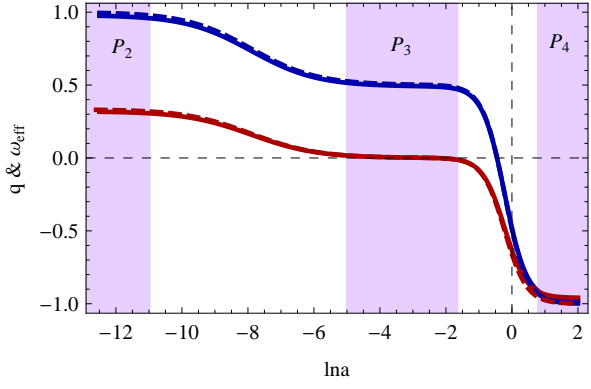}}}}
\vspace{-12pt}{\caption{Behavior of the deceleration parameter $q$ (blue curves) and the effective equation of state parameter $\omega_{\text{eff}}$ (red curves) in $\Lambda$CDM model (dashed curves) and EMSG (solid curves). For EMSG The initial conditions is the same as in Fig. \ref{2}. In the case of $\Lambda$CDM, the initial conditions have been set in such a way to recover the same values for cosmic density parameters as in EMSG at $\ln a=0$. 
}\label{3}}
\end{figure}

Before closing this section we remind that for the whole description presented in this section, we assumed $\eta>0$. However as we mentioned before, there is no a-priori reason why this parameter should be positive. However if $\eta<0$ then using \eqref{fr1} one can conclude that there will not be a bounce in this theory. More importantly using the dynamical system approach we found that there is no stable late time accelerated expansion in this case. It should be notes for $\eta<0$ the autonomous differential equations can be simply obtained from \eqref{n12} by replacing $\Omega_{\text{max}}$ with $\Omega_{\eta}$. Where $\Omega_{\eta}>0$ is
\begin{equation}
\Omega_{{\eta}}=-\frac{\kappa^2}{6\eta H^2}=-\frac{1}{12}{\frac {3{\Omega_{{\text{m}}}}^{2}+14 \Omega_{{\text{r}}}  
\Omega_{{\text{m}}}  +12{\Omega_{{\text{r}}}}^{2}}{-1+\Omega_{{
\Lambda}} \left( t \right) +\Omega_{{\text{m}}}+\Omega_{{\text{r}}}}}
\end{equation}
then finding the fixed points and their stability is completely similar to the $\eta>0$ case. There are three fixed points $(\Omega_{\text{m}}^c,\Omega_{\text{r}}^c,\Omega_{\Lambda}^c)$ as
\begin{equation}
\begin{split}
&P_1'=(0,1,0)\\&
P_2'=(1,0,0)\\&
P_3'=(0,\Omega_{\text{r}},1-2\Omega_{\text{r}})
\end{split}
\end{equation}
$P_1'$ corresponds to an unstable radiation dominated phase, and the corresponding eigenvalues are $(1,12,4)$. $P_2'$ is an unstable matter dominated phase with two positive eigenvalues $(3,9,-1)$. As expected, these points are unstable.

On the other hand, $P_3'$ is not a fixed point but a fixed "line". This line exists provided that $\Omega_{\text{r}}<1/2$. The corresponding eigenvalues are $(0,12,-6/5)$. Surprisingly, all the points on this curve are unstable even the point $(0,0,1)$. This means that the late time de Sitter universe is not stable when $\eta<0$. It is interesting that the correction terms in EMSG, which are expected to be important in early times, substantially disturb the standard paradigm specially in the late times. As a final remark on the cosmology of EMSG, we should stress that these conclusions are true only for a bare cosmological constant. In the presence of the other fields and the vacuum energy density, in principle, the cosmological consequences would be different.
\section{charged black holes in EMSG}
\label{cbh}
As we mentioned before, we expect that in high density regions, for example within a black hole, EMSG to be different form GR. On the other hand, it is obvious that in the vacuum, where the matter energy density is zero, EMSG is equivalent to GR. Consequently, Schwarzschild-de Sitter metric and Kerr metric are also solutions for EMSG field equations. In order to show some interesting differences of these two theories, let us examine a charged black hole. Outside the black hole there exists matter field ($T_{\mu\nu}\neq 0 $), i.e. the electromagnetic field with Lagrangian density $L_m=-\frac{1}{4}F_{\mu\nu}F^{\mu\nu}=-\frac{1}{4}\mathbf{F}^2$, where $F^{\mu\nu}=\partial _{\mu} A_{\nu}-\partial _{\nu} A_{\mu}$ is the electromagnetic field strength tensor. It is important to mention that EMSG not only changes the gravitational theory but somehow postulates a universal modification to all matter field's equations of motion. In the case of the electromagnetic filed, EMSG adds some non-linear terms to Maxwell equations. In this sense, EMSG is reminiscent of the Born-Infeld non-linear electrodynamics \cite{Born:1934gh}. Albeit the field equations are different. More specifically, as we will show, unlike the Born-Infeld theory EMSG does not lead to a non-singular electric field for a point charge. Using the above mentioned Lagrangian for the electrodynamics we have
\begin{equation}
T_{\mu\nu}=F^{\alpha}_{~\mu}F_{\alpha\nu}-\frac{1}{4}g_{\mu\nu}\mathbf{F}^2
\end{equation}
one can simply show that $T=0$. Therefore 
\begin{equation}
\mathbf{\Psi}_{\mu\nu}=-L_m T_{\mu\nu}-2 T^{\alpha\beta}\frac{\partial^2 L_m}{\partial g^{\alpha\beta}g^{\mu\nu}}
\end{equation}
On the other hand 
\begin{equation}
\frac{\partial^2 L_m}{\partial g^{\alpha\beta}g^{\mu\nu}}=-\frac{1}{2}F_{\alpha\mu}F_{\beta\nu}
\end{equation}
Consequently $\mathbf{\Psi}_{\mu\nu}$ takes the following form
\begin{equation}
\begin{split}
\mathbf{\Psi}_{\mu\nu}&=\frac{1}{4}\mathbf{F}^2T_{\mu\nu}+T^{\alpha\beta}F_{\alpha\mu}F_{\beta\nu}\\& =F^{\gamma\alpha}F_{\gamma}^{~\beta}F_{\alpha\mu}F_{\beta\nu}-\frac{1}{16}g_{\mu\nu}(\mathbf{F}^2)^2
\end{split}
\end{equation}
Also after some algebra we find
\begin{equation}
\begin{split}
&T_{\mu}^{~\sigma}T_{\nu\sigma}=F^{\gamma\alpha}F_{\gamma}^{~\beta}F_{\alpha\mu}F_{\beta\nu}+\frac{1}{16}g_{\mu\nu}(\mathbf{F}^2)^2\\&
-\frac{1}{4}g_{\mu\nu}\mathbf{T}^2=-\frac{1}{4}g_{\mu\nu}F^{\alpha}_{~\theta}F_{\alpha\rho}F^{\gamma\theta}F_{\gamma}^{~\rho}+\frac{1}{16}g_{\mu\nu}(\mathbf{F}^2)^2
\end{split}
\end{equation}
Therefore the field equations \eqref{fe1} can be rewritten as
\begin{equation}
\begin{split}
G_{\mu\nu}+&\Lambda g_{\mu\nu}=\kappa\Big(F^{\alpha}_{~\mu}F_{\alpha\nu}-\frac{1}{4}g_{\mu\nu}\mathbf{F}^2\Big)\\&+2\eta\Big(\frac{1}{16}g_{\mu\nu}(\mathbf{F}^2)^2+2F^{\gamma\alpha}F_{\gamma}^{~\beta}F_{\alpha\mu}F_{\beta\nu}\\&-\frac{1}{4}g_{\mu\nu}F^{\alpha}_{~\theta}F_{\alpha\rho}F^{\gamma\theta}F_{\gamma}^{~\rho}\Big)
\end{split}
\label{n6}
\end{equation}
Although these field equations seem very complicated, they can be solved analytically for a charged black hole.

Now we need the generalized form of the Maxwell equations. we recall again that EMSG not only changes the gravitational filed equations but also modifies the matter field equations in the high curvature regime. To find the modified electrodynamic field equations in the vacuum, we need to vary the following part of the action \eqref{action} with respect to $A_{\mu}$
\begin{equation}
\int \sqrt{-g}(\frac{1}{4}\mathbf{F}^2+\frac{\eta}{2\kappa}\mathbf{T}^2)d^4x
\end{equation}
Variation leads to the following Euler-Lagrange equation
\begin{equation}
\nabla_{\mu}\frac{\partial(\kappa\mathbf{F}^2+2\eta\mathbf{T}^2) }{\partial(\nabla_{\mu}A_{\nu})}=0
\end{equation}
Note that $\frac{\partial(\kappa\mathbf{F}^2+2\eta\mathbf{T}^2)}{\partial A_{\nu}}=0$. Also we remind that
\begin{equation}
\frac{\partial F_{\alpha\beta}}{\partial(\nabla_{\mu}A_{\nu})}=\delta^{\mu}_{\alpha}\delta^{\nu}_{\beta}-\delta^{\mu}_{\beta}\delta^{\nu}_{\alpha}
\end{equation}
Using this equation it is straightforward to obtain
\begin{equation}
\begin{split}
&\frac{\partial \mathbf{F}^2}{\partial(\nabla_{\mu}A_{\nu})}=4F^{\mu\nu}\\&
\frac{\partial \mathbf{T}^2}{\partial(\nabla_{\mu}A_{\nu})}=8F_{\gamma}^{~\nu}F^{\mu\rho}F^{\gamma}_{~\rho}-2 F^{\mu\nu}\mathbf{F}^2
\end{split}
\end{equation}
Finally the modified version of the Maxwell equations in the vacuum are
\begin{equation}
\nabla_{\mu}F^{\mu\nu}=\frac{\eta}{\kappa}\nabla_{\mu}\left[4F_{\gamma}^{~\nu}F^{\mu\rho}F^{\gamma}_{~\rho}-F^{\mu\nu}\mathbf{F}^2\right]
\label{maxwell}
\end{equation}
\begin{equation}
\nabla_{[\mu}F_{\nu\lambda]}=0
\label{max2}
\end{equation}
Equations \eqref{max2} are geometrical equations valid independently of the Lagrangian chosen. Equations \eqref{maxwell} and \eqref{max2} are coupled to equations \eqref{n6}. In fact $F_{\mu\nu}$ enters the gravitational field equations \eqref{n6} and
the metric $g_{\mu\nu}$ enters the electromagnetic field equations \eqref{maxwell} through the covariant derivative. These equations, i.e. \eqref{n6}, \eqref{maxwell} and \eqref{max2}, form a complete set of differential equations to obtain the metric components and the electromagnetic fields. From equations \eqref{maxwell} it is clear that the correction term is proportional to $\rho_{\text{max}}^{-1}$. Therefore, as expected, in low energy regimes there would be no deviations from standard electrodynamics. Now, let us start with a spherically symmetric space-time as 
\begin{equation}
ds^2=-\phi(r)f(r)dt^2+\frac{dr^2}{f(r)}+r^2d\Omega^2
\end{equation}
As in GR, regarding the spherical symmetry of the metric, one may use the following components for $F_{\mu\nu}$: $F_{tr}=-F_{rt}=E(r)$, where $E(r)$ is the electric field, and the other components are all zero. With this choice, the equations \eqref{max2} are automatically satisfied. Also from the field equations, it turns out straightforwardly that $\phi(r)=1$. On the other hand equation \eqref{maxwell} has only one nonzero component as
\begin{equation}
\Big(1+\frac{6\eta}{\kappa}E(r)^2\Big)\frac{dE(r)}{dr}+\frac{2E(r)}{r}+\frac{4\eta}{\kappa r}E(r)^3=0
\label{n14}
\end{equation}
Fortunately this equation can be integrated and the solution is
\begin{equation}
 E(x)=\sqrt{\frac{\kappa}{6\eta}}\left(\frac{x^{-1/3}\left(1+\sqrt{1+x^4}\right)^{2/3}-x}{x^{1/3}\left(1+\sqrt{1+x^4}\right)^{1/3}}\right)
 \label{el}
\end{equation}
where the new parameter $x$ is defined as $r=(27 q^2\eta/2\kappa)^{1/4}x$, and $q$ is the integration constant. As we will see $q$ is related to the total electric charge of the black hole.

On the other hand, the field equations \eqref{n6} lead to two independent differential equations. More specifically, $tt$ and $rr$ components lead to a same equation and $\theta\theta$ and $\phi\phi$ components lead to another independent equation. $tt$ and $\theta\theta$ components are respectively
\begin{equation}
r\frac{df(r)}{dr}-1+f(r)+\frac{1}{2}\kappa r^2 E(r)^2\Big(1+\frac{3\eta}{\kappa}E(r)^2\Big)=0
\label{n7}
\end{equation}
\begin{equation}
r\frac{df(r)}{dr}+\frac{r^2}{2}\frac{d^2f(r)}{dr^2}-\frac{1}{2}\kappa r^2 E(r)^2\Big(1+\frac{\eta}{\kappa}E(r)^2\Big)=0
\label{n8}
\end{equation}
At the first glance it seems that we have two differential equations for one unknown function $f(r)$. However, one may easily show that equation \eqref{n8} is not also independent equation. In fact differentiating \eqref{n7} with respect to $r$ and combining with \eqref{n14}, we get equation \eqref{n8}. Hence we need to solve only equation \eqref{n7}. The solution can be expressed as
 \begin{equation}
f(r)=1-\frac{\kappa M}{4\pi r}-\frac{\Lambda r^2}{3}-\frac{\kappa}{2r}\int \left(E^2(r)+\frac{3\eta}{\kappa}E^4(r)\right)r^2 dr
 \end{equation}
where $M$ is the "mass". Note that the electric field is singular at $r=0$. One may easily verify that for small $\eta/\kappa$ the Reissner-Nordstrom metric is recovered. In fact the expansion of $f(r)$ and the electric field $E(r)$ to the first order of $\eta/\kappa$ is 
\begin{equation}
\begin{split}
&f(r)\simeq 1-\frac{\Lambda r^2}{3}-\kappa\left(\frac{M}{4\pi r}-\frac{q^2}{2r^2}+\frac{q^4}{10r^6}\frac{\eta}{\kappa}\right) \\&
E(r)\simeq\frac{q}{r^2}-\frac{2 q^3}{r^6}\frac{\eta}{\kappa}
\end{split}
\label{app}
\end{equation}
The metric is still singular at $r=0$. However, this singularity is inside the horizon. Since this solution is an exterior solution to a charged black hole, the $r=0$ singularity does not imply a singular solution. One needs to find interior solutions in the presence of normal matter in order to check the singularity structure of EMSG. For example the singularity-free expanding/collapsing FRW universe that we have already explored in this paper lies in this category. Therefore if a finite maximum density arises in the interior solutions, more specifically in the gravitational collapse, then EMSG propose an entirely singularity-free universe. 

\section{conclusion}
\label{conc}
In this paper a new covariant generalization of GR is developed. This theory allows the existence of a term proportional to $T_{\alpha\beta}T^{\alpha\beta}$ in the action. Therefore we referred to this theory as Energy-Momentum Squared Gravity (EMSG). EMSG is different from GR only in the presence of matter sources. In this theory the correction term can be defined only when the Lagrangian density for the matter content is specified. Therefore in order to find the field equations, one must first vary the matter action with respect to the gravitational degrees of freedom. Although this feature is not the case in GR, it appears in theories which introduce correction terms including the energy-momentum tensor in the generic action

 Applying this theory to a homogeneous and isotropic space-time, we find that there is a maximum energy density $\rho_{\text{max}}$, and correspondingly a minimum length scale $a_{\text{min}}$, at early universe. In other words, we showed that there is a bounce at early times and consequently the early time singularity is avoided. We found the exact value of $\rho_{\text{max}}$. Also we estimated the minimum value of the cosmic scale factor. Moreover, the dynamical system method has been used to investigate the cosmological behavior of EMSG. It turned out that EMSG possesses a true sequence of cosmological eras (or fixed points). Comparing to $\Lambda$CDM model, there is an extra duty for the cosmological constant in this theory. In fact, a positive $\Lambda$ is necessary for the existence of a regular bounce at early universe.

Also an exact solution for a charged black hole has been found. We recall that Schwarzschild and Kerr metrics are also solutions for EMSG field equations. However, the charged black hole solution in EMSG is different from the standard Reissner-Nordstr\"{o}m space-time.

As a further study it is needed to check the existence of stable compact stars in EMSG; for such a study in the context of EiBI see \cite{compact}. It is also necessary to investigate the consequences of the rapid decrease of $\Omega_{\text{r}}$ and the accelerated expansion right after the bounce. Finally one may expect quantum effects to become important at ultra-short distances and ultra-high energy densities. Although in order to avoid these effects one may require $\rho_{\text{max}} < \rho_{\text{p}}$ and $a_{\text{min}}> l_{\text{p}}$ where $\rho_{\text{p}}$ is the Planck density and $l_{\text{p}}$ the Plancklength. Using the current value of the radiation energy density and scaling $a_0=1$, one can easily show that if $\eta> \hbar G^3$ then both conditions are satisfied. If this constraint is consistent with the cosmological observations, then, in EMSG, the universe may not enter a quantum era during its evolution.

\section{Acknowledgments}
We thank the anonymous referee for useful and constructive comments. M. Roshan is grateful to Ahmad Ghodsi and Luca Amendola for valuable discussions, and to Salvatore Capozziello for valuable comments. F. Shojai is grateful to University of Tehran for supporting this work under a grant provided by the university research council.

\end{document}